\documentclass[preprint2]{aastex6}
\usepackage{floatrow}
\usepackage[caption=false]{subfig}
\newcommand{\aver}[1]{\ensuremath{\left\langle#1\right\rangle}}

\begin{document}
\title{Particle Acceleration and Heating by Turbulent Reconnection}
\author{Loukas Vlahos, Theophilos Pisokas, Heinz Isliker, Vassilis Tsiolis}
\affil{Department of Physics,
 Aristotle University of Thessaloniki\\
GR-52124 Thessaloniki, Greece
   }

\and

\author{Anastasios Anastasiadis}
\affil{Institute for Astronomy, Astrophysics, Space Applications and Remote Sensing,\\
 National Observatory of Athens\\
 GR-15236 Penteli, Greece
   }
\date{\today}
\begin{abstract}
Turbulent flows in the solar wind, large scale current sheets, multiple current sheets, and shock waves lead to the formation of environments in which a dense network of current sheets is established and sustains  ``turbulent reconnection''.
We constructed a 2D grid on which a number of randomly chosen grid points are acting as {\bf scatterers} (i.e.\ magnetic clouds or current sheets). Our goal is to examine how test particles respond inside this {\bf large scale} collection of scatterers.  We study the energy gain of individual particles, the evolution of their energy distribution and  their escape time distribution. We have developed a new method to estimate the transport coefficients  from the dynamics of the interaction of the particles with the scatterers. Replacing the ``magnetic clouds''  with current sheets, we have proven that the energization  processes can be more efficient 
depending on the strength of the effective electric fields inside the current sheets and their statistical properties.
 Using the estimated transport coefficients and solving the Fokker-Planck (FP) equation we can recover the energy distribution of the particles only for the sstochastic Fermi process. We have shown that the evolution of the particles inside a turbulent reconnecting volume is not a solution of the FP equation, since the interaction of the particles with the current sheets is ``anomalous'', in contrast to the case of the second order Fermi process.
\end{abstract}
%\maketitle

\section{Introduction}

\cite{Fermi49} introduced a fundamental stochastic process to solve the problem of particle energization (heating and/or acceleration) in space and astrophysical plasmas. His goal was to resolve the mystery of the stable energy distribution of Cosmic Rays (CR) \citep[see details in][]{Longair11}. The core of his idea had a larger impact on non-linear processes in general and has been the driving force behind all subsequent theories on charged particle energization. He assumed that high energy particles with speed close to the speed of light collide with {\b magnetic clouds} which move in random directions with speed $V$ close to the local  Alfv\'en speed. The reflections of the charged particles at the magnetic clouds, heat or accelerate the particles to substantial energies. The rate of the energy gain for the charged particles is proportional to the square of the ratio of the magnetic cloud speed to the speed of light $(V/c)^2$.  A more realistic proposal was put forward initially by  \cite{Kulsrud71}. The magnetic clouds were replaced by a Kolmogoroff spectrum of {\bf low amplitude MHD waves} and the energization processes was called {\bf ``stochastic heating and acceleration by (weak) turbulence''}.

Research on reconnecting magnetic fields has undergone a dramatic evolution recently due mostly to the development of the numerical simulation techniques. Long current sheets or multiple interacting current sheets will form, on a short time scale, a turbulent environment, consisting of a collection of current sheets \citep{Matthaeus86,Galsgaard96,Drake06,Onofri06}, \citep[see also the recent reviews][]{Cargill12,Lazarian12}. On the other hand, Alfv\'en waves and large scale disturbances traveling along complex magnetic topologies will drive magnetic discontinuities by reinforcing existing current sheets or form new unstable current sheets \citep[see][]{Biskamp89,Lazarian99,Dmitruk04,Arzner04}. 

The goals of this article are to introduce
three new and important elements
in the current discussion of turbulent reconnection in
{\bf large scale systems}: (a) the study of the
characteristics of the energy gain of individual particles, (b)  the use of the same framework of global and statistical analysis for two types of scatterers, (i) magnetic clouds, which are representative of stochastic energy gain, (ii) {\bf Unstable Current Sheets (UCS)}, which are representative of systematic energy gain,  (c) the development of a  new method to estimate the {\bf transport coefficients from the dynamics of the interaction of the particles with the scatterers}. %-----------------------------------------------------------------------------------
%-----------------------------------------------------------------------------------
\section{Fermi type energization of particles}

\cite{Fermi49} based his estimates for the proposed acceleration mechanism on several assumptions \citep[see][]{Longair11}.
The particles move with relativistic velocity $u$ and the scatterers (``magnetic clouds'') move with mean speed $V$ much smaller than the speed of light. The energy gain or loss of the particles interacting with the scatterers is
\begin{equation}\label{energyF}
	\frac{\Delta W}{W}\approx\frac{2}{c^2}(V^2-\vec{V} \cdot \vec{u}) ,
\end{equation}
where for head on collisions $\vec V\cdot \vec u < 0$ and the particles gain energy, for overtaking collisions
$\vec V\cdot \vec u > 0$ and the particles lose energy. The rate of energy gain in Eq. (1) includes both,
a first and a second order term. For relativistic particles the first order term dominates the
energy gain. For non-relativistic particles both terms are second order.

The rate of energy gain for relativistic particles is estimated as
$dW/dt=W/t_{acc},$
where $t_{acc}=(3\lambda c)/(4V^2)$
and $\lambda$ is the mean free path the particles travel between the scatterers. Assuming that the distribution of the scatterers is uniform inside the acceleration volume and their density is $n_{sc}$, the mean free path will be
	$\lambda \approx  (\sqrt[3]{n_{sc}})^{-1}.$
The particles are not trapped inside the scatterers, {\bf their interaction is instantaneous} and the temporal evolution of the mean energy is
\begin{equation}\label{Energy}
	\aver{W(t)} = W_0e^{t/t_{acc}} .
\end{equation}
\cite{Fermi49} used the FP equation in order to estimate the change of the energy distribution $n(W,t)$ of the accelerated particles. In order to simplify the diffusion equation, he assumed that spatial diffusion is not important and the particles diffuse only in energy space,
\begin{equation} \label{diff}
\frac{\partial n}{\partial t}+\frac{\partial }{\partial W} \left [ F n -\frac{\partial [D n]  }{\partial W} \right ]=
	-\frac{n}{t_{esc}}+Q  ,
\end{equation}
where $t_{esc}$ is the escape time from an acceleration region with characteristic length $L$, $Q$ is the injection rate,
$D$ is the energy diffusion coefficient
\begin{equation}
	D(W,t) =\frac{\aver{\left(W(t+\Delta t)- W(t)\right)^2}_W}{2\Delta t},
	\label{eq:DWW}
\end{equation}
and
\begin{equation}
	F(W,t) =\frac{\aver{W(t+\Delta t)- W(t)}_W}{\Delta t},
	\label{eq:FW}
\end{equation}
is the energy convection coefficient representing the systematic
acceleration, which, as mentioned, here takes the form
$F(W,t)=W/t_{acc}$.
With $\aver{...}_W$ we denote the conditional average that $W(t)=W$
(see e.g.\ \cite{Ragwitz2001}).
Fermi reached his famous result by assuming that: (a) the particles reach a steady state before escaping from the acceleration volume and (b) the energy diffusion coefficient approaches zero asymptotically for the relativistic particles and the acceleration is mainly due to the systematic acceleration term ($F$).
Based on these assumptions, the stationary solution of Eq.~(\ref{diff}) simply is $n(W)\sim W^{-k}$,
where
$k = 1+t_{acc}/t_{esc}.$
The index $k$ approaches 2 (which is close to the observed value for the CR) only if $t_{acc} \approx t_{esc}.$ In most recent theoretical studies of the second order Fermi acceleration the escape time (which is so crucial for the estimate of $k$) is difficult  to estimate quantitatively.

%-----------------------------------------------------------------------------------

 We will  expand the initial Fermi model in this article,  by replacing  the scatterers by randomly distributed UCS, which represents the environment present in turbulent reconnection in a fragmented large scale system. 
 In several recent articles the 3D evolution and the fragmented  UCS
 %internally
 has been analysed (see \cite{Guo15, Dahlin15}), using Particle in Cell numerical codes, and it has been found that the curvature drift competes with the electric field in the efficiency of particle acceleration inside the UCS.
It will be a natural continuation of the work presented here to study also the curvature
drift case, 
here we focus
%for simplicity
on the acceleration by the electric fields. 
%The interaction of particles with models of UCS has been studied extensively,
%  and  
The particle dynamics inside the UCS is complex since internally the UCS are also fragmented and the particles that interact with the fragments of the UCS can lose and gain energy on the microscopic level of description.  
Yet, on the average and over the entire simulation domain, the particles gain energy systematically before exiting the UCS, see 
Fig.\ 6(c) of \cite{Guo15} and the related discussion. 
The energy gain  is a weak function of energy in the
case of electric field acceleration and proportional to the energy in the case of curvature
drift.
%Therefore the particle interaction with a UCS is always positive but
In this article we estimate the {\bf macroscopic} energy gain by the simple formula
 \begin{equation}\label{e:dW_cs}
	\Delta W = |q| E_\textrm{eff}\ \ell_\textrm{eff}  ,
\end{equation}
 where $E_\textrm{eff} \approx  (V/c) \; \delta B$ is the measure of the effective electric field of the UCS, and $\delta B$ is the fluctuating magnetic field encountered by the particle, which is of stochastic nature, as related to the stochastic fluctuations induced by  reconnection.  $\ell_\textrm{eff}$ is the characteristic length of the interaction of the particle with the UCS and should be proportional to $E_{eff}$, since small $E_{eff}$ will be related to small scale  UCS. 
 The scenario of the method used here is: particles  approach the scatterers  with an initial energy $W_0$ and depart with a energy $W=W_0+ \Delta W$, where $\Delta W$ {\bf on the macroscopic level}  always is positive and follows the statistical properties of the fluctuations $\delta B$.

%-----------------------------------------------------------------------------------
%-----------------------------------------------------------------------------------

%-----------------------------------------------------------------------------------
\section{A Fermi lattice gas model for turbulent reconnection}
We constructed a 2D grid $(N \times N),$  with linear size $L$. Each grid point is set as either \emph{active} or \emph{inactive}, i.e.~scatterer or not. Only a small fraction R $(1-15\%)$ of the grid points are active.  The mean free path of the particles moving inside the grid with minimum distance $\ell=L/(N-1)$ is $\lambda_{sc}=\ell/R.$ When a particle encounters an active grid point it is renewing its energy state depending on the physical characteristic of the scatterer (magnetic cloud or UCS).

At time, $t=0$ all particles are located at random positions on the grid. The injected distribution $n(W, t=0)$ is Maxwellian with temperature $T$. The initial direction of motion of every particle is selected randomly. The particles' individual time $t_i$ is also adjusted between scatterings as $t_{i+1}=t_{i}+ \Delta t, \;\; \Delta t= l_i/u_i,$ with $u_i$ the particle velocity and $l_i$ the distance the particle travels between scatterings. The particles move in a random direction after  interaction with the scatterers, being always confined to follow the grid-lines.
It is to
note that the consequent large angle scattering takes place in position space, and not in
velocity space, the large angle scattering is unrelated with the particle energy, and its role
is to implement a spatial random walk process on a grid that basically is influencing only
the timing of the energization process.
We mainly consider electrons and will just briefly comment on the energization of ions.

%-----------------------------------------------------------------------------------
\paragraph{Random ``scattering'' by magnetic clouds }
 We start our analysis using the standard stochastic Fermi accelerator,
Eq.\ (\ref{energyF}), in order to validate our method for the estimate
of the transport coefficients and the solution of the Fokker Planck equation, since this
accelerator has been already discussed in the literature using many different approaches.
The parameters used in this article are related to the plasma parameters in the low solar corona. We choose the strength of the magnetic field to be $B=100\ G,$ the density of the plasma  $n_0= 10^9\ cm^{-3}$ and the ambient temperature around $10\ eV.$ The Alfv\'{e}n speed is  $V_A \approx 7\times 10^8\ cm/sec$, so $V_A$ is comparable with the thermal speed of the electrons. The energy increment is  $(\Delta W/W) \sim (V_A/c)^2\approx 5 \times 10^{-4}$ and the length of the simulation box is $10^{10}\ cm$.  We consider an open grid, so particles escape from the  accelerator when they reach any boundary of the grid, at $t_i = t_{esc,i}.$  We assume in this set-up that only $R=10\%$ of the  $601\times 601$ grid points are active.

\begin{figure}[ht]
\sidesubfloat[]{\includegraphics[width=0.45\columnwidth]{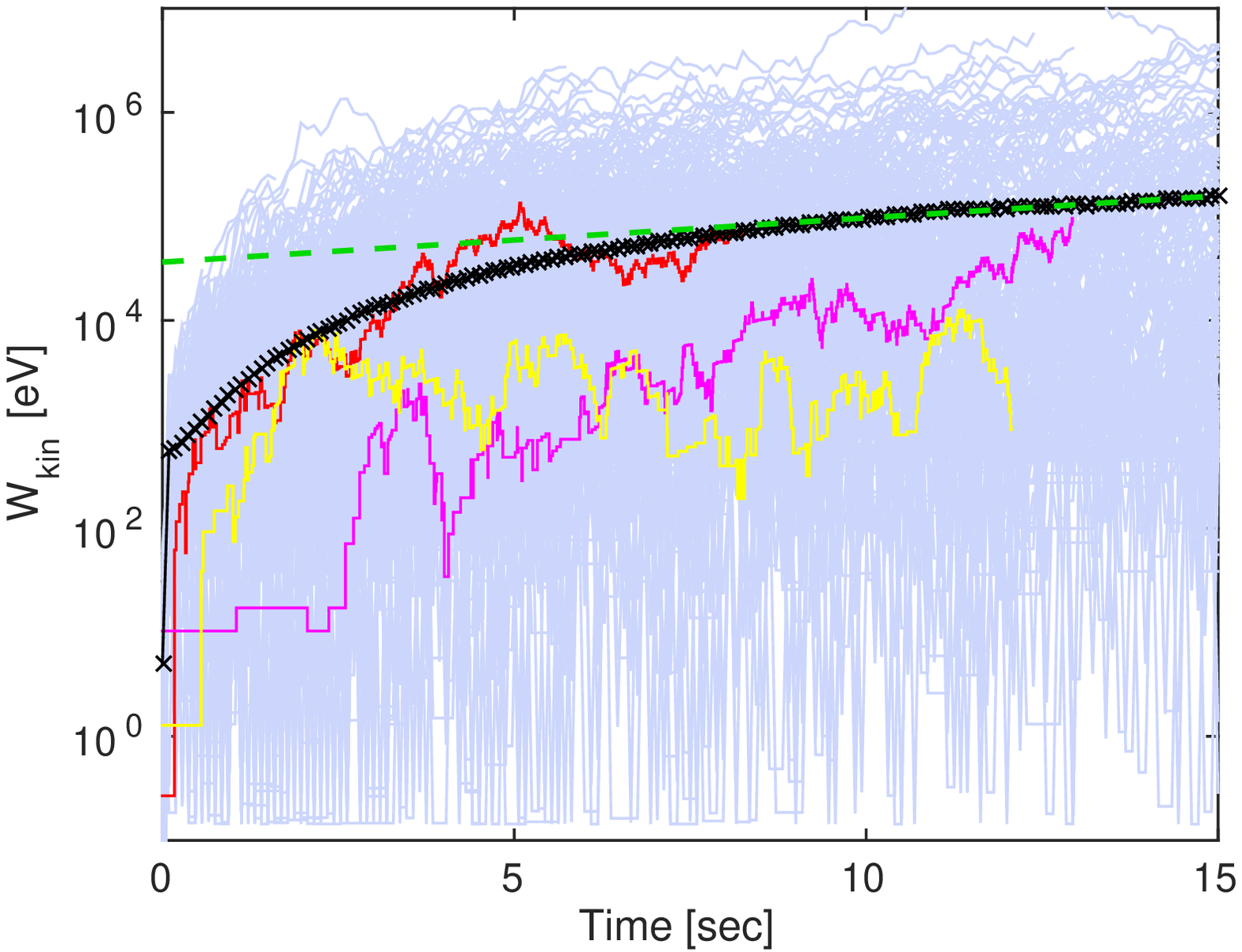}%
		\label{f:distr_sof:mW}}
		\sidesubfloat[]{\includegraphics[width=0.45\columnwidth]{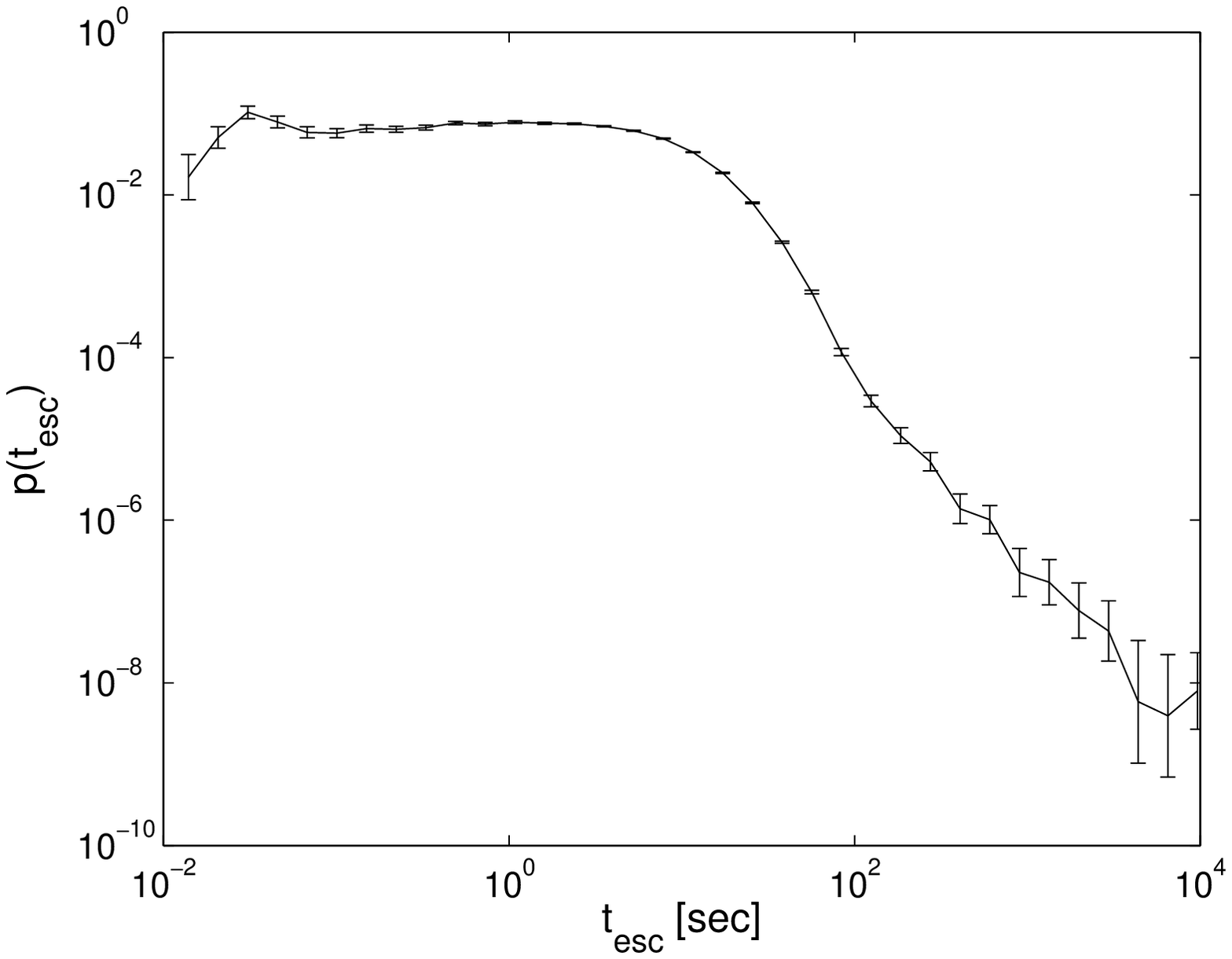}%
		\label{f:distr_sof:tdistr}}\hfill\\
	\sidesubfloat[]{\includegraphics[width=0.45\columnwidth]{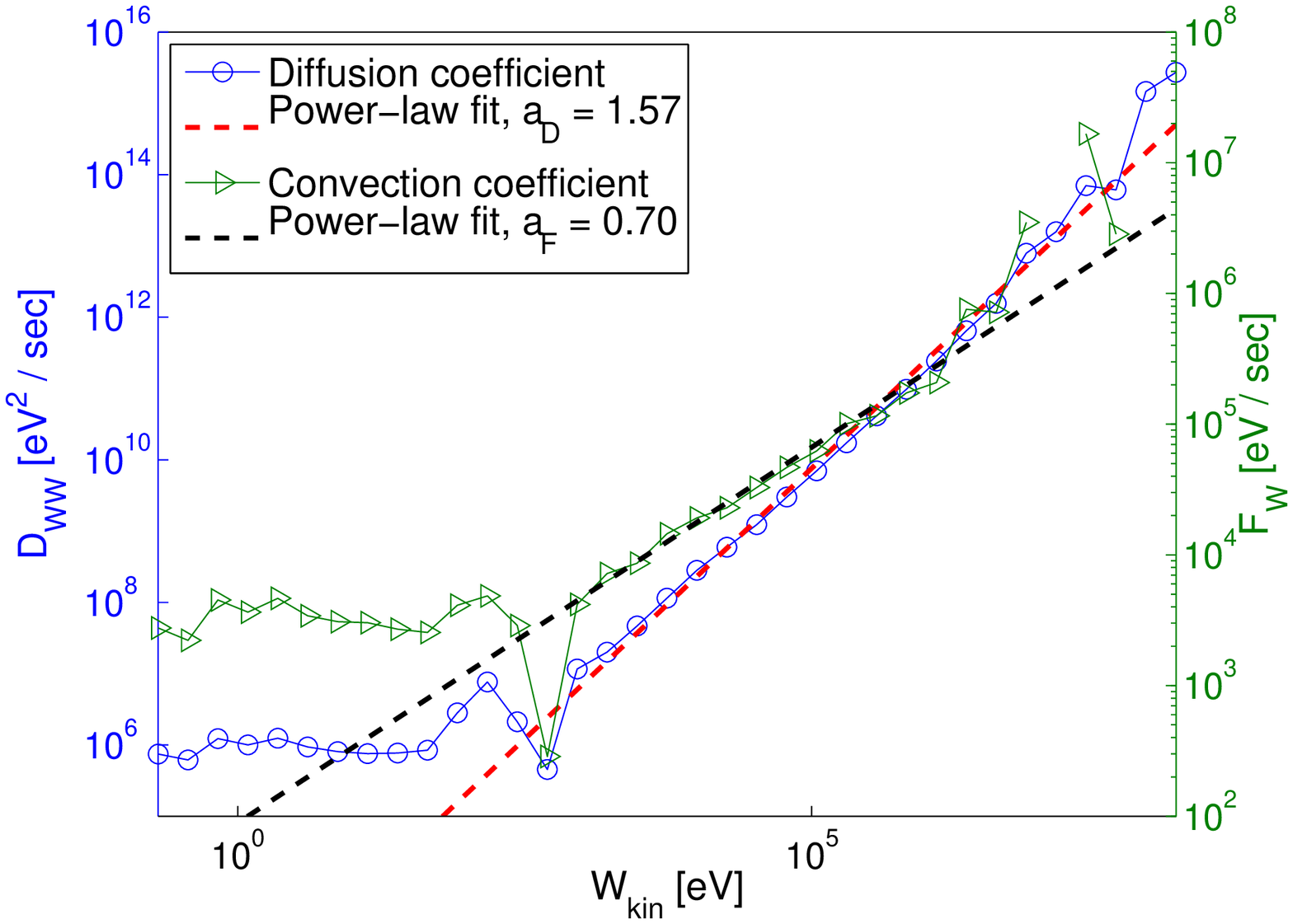}%
		\label{f:distr_sof:DF_W}}
		\sidesubfloat[]{\includegraphics[width=0.45\columnwidth]{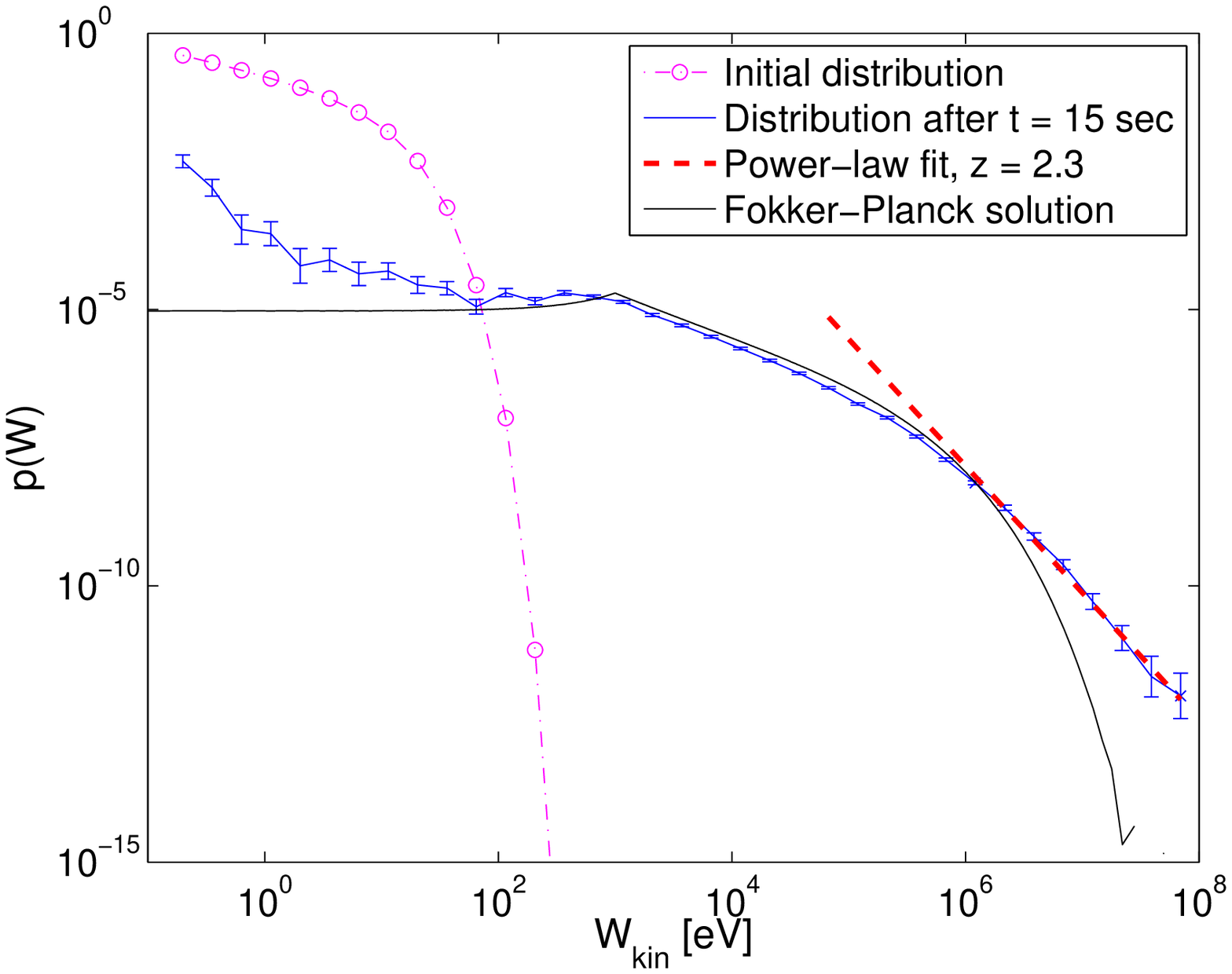}%
		\label{f:distr_sof:Wdistr}}
	\caption{\textit{\protect\subref{f:distr_sof:mW} Mean energy increase as a function of time and the energy evolution of typical particles. \protect\subref{f:distr_sof:tdistr} The escape time distribution of particles.  \protect\subref{f:distr_sof:DF_W} The energy diffusion and convection coefficients  as functions of the kinetic energy. \protect\subref{f:distr_sof:Wdistr} Energy distribution at $t=0$ and $t=15\ sec$ for particles remaining inside the box, together with the solution of the FP equation at final time.
	%The initial conditions used in this run have been outlined in the text.
			}}\label{f:distr_sof}
\end{figure}

The temporal evolution of the mean kinetic energy of the particles and the kinetic energy evolution of typical particles  are shown in Fig.~(\ref{f:distr_sof:mW}). The motion of the particles is typical for a stochastic system with random-walk like gain and loss of energy before exiting the simulation box. The mean energy increases exponentially (after a brief initial period of a few  seconds), as is expected from the analysis presented by Fermi (see Eq.~(\ref{Energy})). The mean free path
is given as  $\lambda_{sc}=\ell/R \approx 1.67\times 10^8\ cm$,
and, using the analytical expression derived by Fermi,  we find  $t_{acc_{th}}=(3\lambda_{sc} c)/(4V_A^2) \approx 8\ sec$. We can also estimate the acceleration time from our simulation (see Fig.~(\ref{f:distr_sof:mW})), by fitting the asymptotic exponential form to the mean kinetic energy, as predicted by Eq.~(\ref{Energy}), which
yields
$t_{acc_{num}}\approx 10\ sec$, a value close to the analytically determined one.  Fig.~(\ref{f:distr_sof:tdistr}) presents the escape time, which is different
for each particle, and we use the median value $(\approx 8\ sec)$ as
an estimate of a characteristic escape time.
In Fig. (\ref{f:distr_sof:Wdistr}) we show the energy distribution function of the particles remaining inside the box after 15 sec. The distribution is a synthesis of a hot plasma  and a power law tail, which is extended to  $100 MeV$, with slope $k \approx 2.3$. If we use the estimates of $t_{acc}$ and $t_{esc}$ reported, we can estimate the index of the power law tail $k=1+t_{acc}/t_{esc} \approx 2.3.$  So the slope of the accelerated particles agrees with the estimates provided by the theory of the stochastic Fermi process.

In Fig.~(\ref{f:distr_sof:DF_W}), the diffusion and convection coefficients at
$ t = 15\ sec$, as functions of the energy, are presented.  The estimate of the coefficients is based on Eqs.\ (\ref{eq:DWW}) and  (\ref{eq:FW}), with $\Delta t$ small, whereto we monitor the energy
of the particles at a number of regularly spaced monitoring times $t^{(M)}_k$,
$k=0,1,...,K$, with $K$ typically chosen as $200$, and we use
$t=t^{(M)}_{K-1}$, $\Delta t = t^{(M)}_{K} - t^{(M)}_{K-1}$ in the
estimates. Also, in order to account for the conditional averaging in
Eqs.\ (\ref{eq:DWW}) and  (\ref{eq:FW}), we divide the energies
$W\left(t^{(M)}_{K-1}\right)_i$ of the particles into a number of logarithmically
equi-spaced bins and perform the requested averages separately
for the particles in each bin.
As Fig.~(\ref{f:distr_sof:DF_W}) shows,
both transport coefficients exhibit a power-law shape, with indexes $a_D = 1.57$ and $a_F = 0.70$, for energies above $1\ keV$,
$
F(W) = A W^{0.70}, \ \ \ \ \ D(W) = B W^{1.57}.$
These estimates clearly depart from the assumptions made  initially by Fermi.

In order to verify the estimates of the transport coefficients,
we insert them in the form of the fit into the FP equation (Eq.\ (\ref{diff})) and
solve the FP equation numerically (including the escape term, and with $Q=0$).
For the integration of the FP equation on the semi-infinite
energy interval $[0,\infty)$, we use the pseudospectral
method, based on the expansion in terms of rational Chebyshev polynomials
in energy space, combined with the implicit backward Euler method
for the time-stepping (see e.g.\ \cite{Boyd2001}). The resulting energy distribution at final time
is also shown in Fig.~(\ref{f:distr_sof:Wdistr}), and it turns out to
coincide very well with the distribution from the particle simulation
in the intermediate energy range that corresponds to the heating of the
population, the power-law tail can though not be reproduced by the
FP solution. The differences below energies of about $10 \; eV$ are of less importance
and can most likely be attributed to the fact that for simplicity we just assumed the transport coefficients to be constant at low energies.

Varying the density of the scatterers in a parametric
study in the range $0.01 < R < 0.2$
and keeping the characteristic length of the acceleration volume constant,
we find that the main characteristics of the distribution remain the same but the heating and the slope of the accelerated particles vary.

 The ions in the asymptotic stage do not appear to have significant differences from the evolution of the electrons.  We can then conclude that stochastic  Fermi processes can heat and accelerate both ions and electrons in the solar corona, yet on different time scales.

%-----------------------------------------------------------------------------------
%-----------------------------------------------------------------------------------
 %-----------------------------------------------------------------------------------
  \paragraph{A model for turbulent reconnection}
  We now use the lattice gas model to estimate the heating and
  acceleration of particles inside a large scale turbulent
  reconnection environment, where a fragmented distribution of UCS
  is present. The setup is $R=0.1$, $N=601$, $V=V_A$ and
  the simulation box has length $10^8\ cm$ and is open. 
 The energy change of a particle that encounters an UCS is now given by
  Eq.~(\ref{e:dW_cs}), and we assume that $\delta B$ takes random values 
  following a power-law distribution with index $5/3$
  (Kolmogorov spectrum), and $\delta B \in [10^{-5} G, 100G].$ 
  We also assume the effective length $\ell_\textrm{eff}$ to be a linear function of $E_{eff}$, $\ell_\textrm{eff} =a E_{eff}+b$, and by restricting the size of $\ell_\textrm{eff}$ to $\ell_{eff} \in [10^3 cm, 10^5 cm]$,  we determine the constants $a,b.$
  Combining all the above we find that the effective electric filed lies  approximately in 
   $E_\textrm{eff} \in [10^{-7} E_D, 
  E_D]$,  where  $E_D$ is the Dreicer field, $E_D\approx 1.6\cdot
  10^{-7}\ statV/cm$.

  \begin{figure}[ht]
  \sidesubfloat[]{\includegraphics[width=0.43\columnwidth]{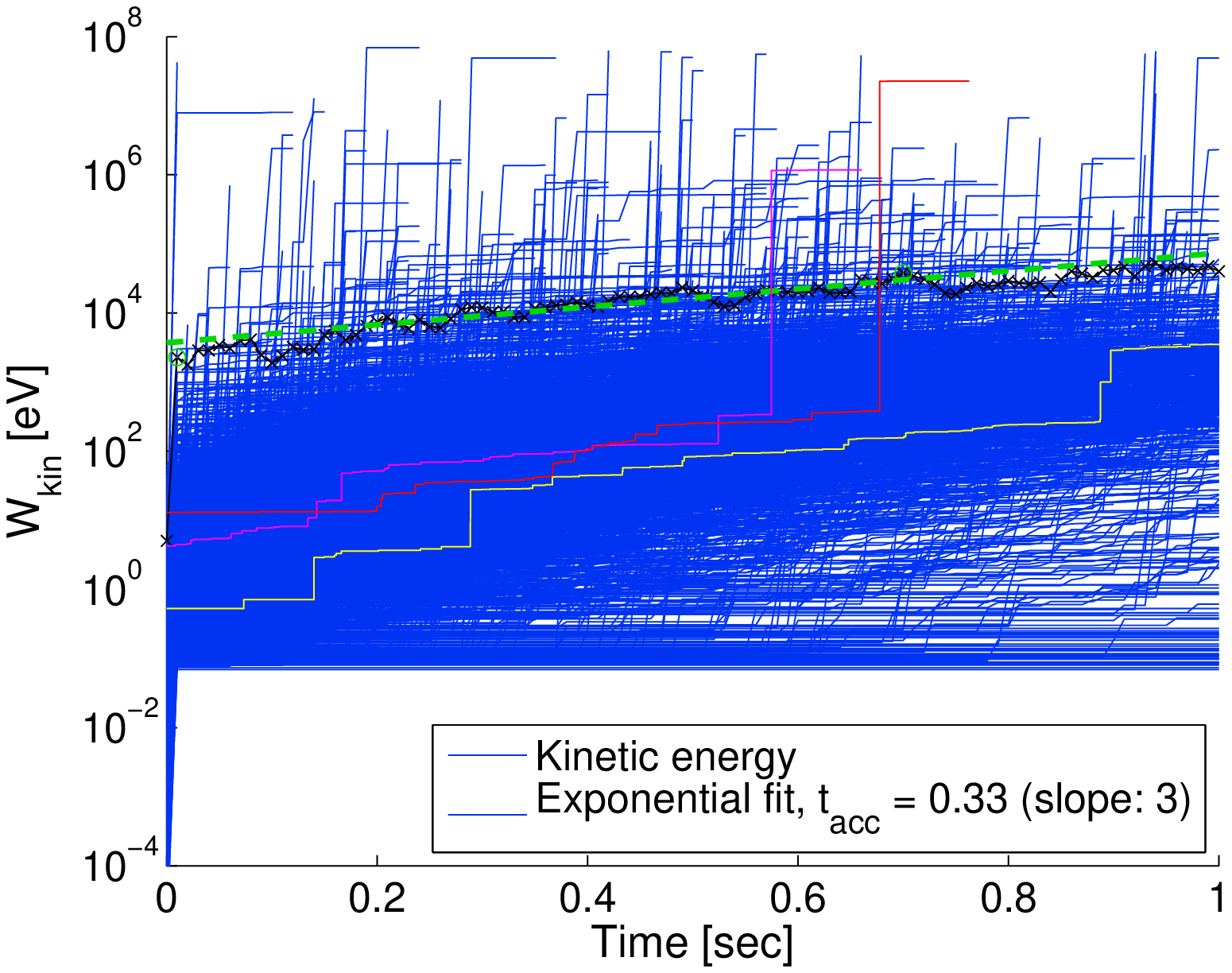}%
                  \label{f:distr_cs:mW}}\hfill%
  \sidesubfloat[]{\includegraphics[width=0.43\columnwidth]{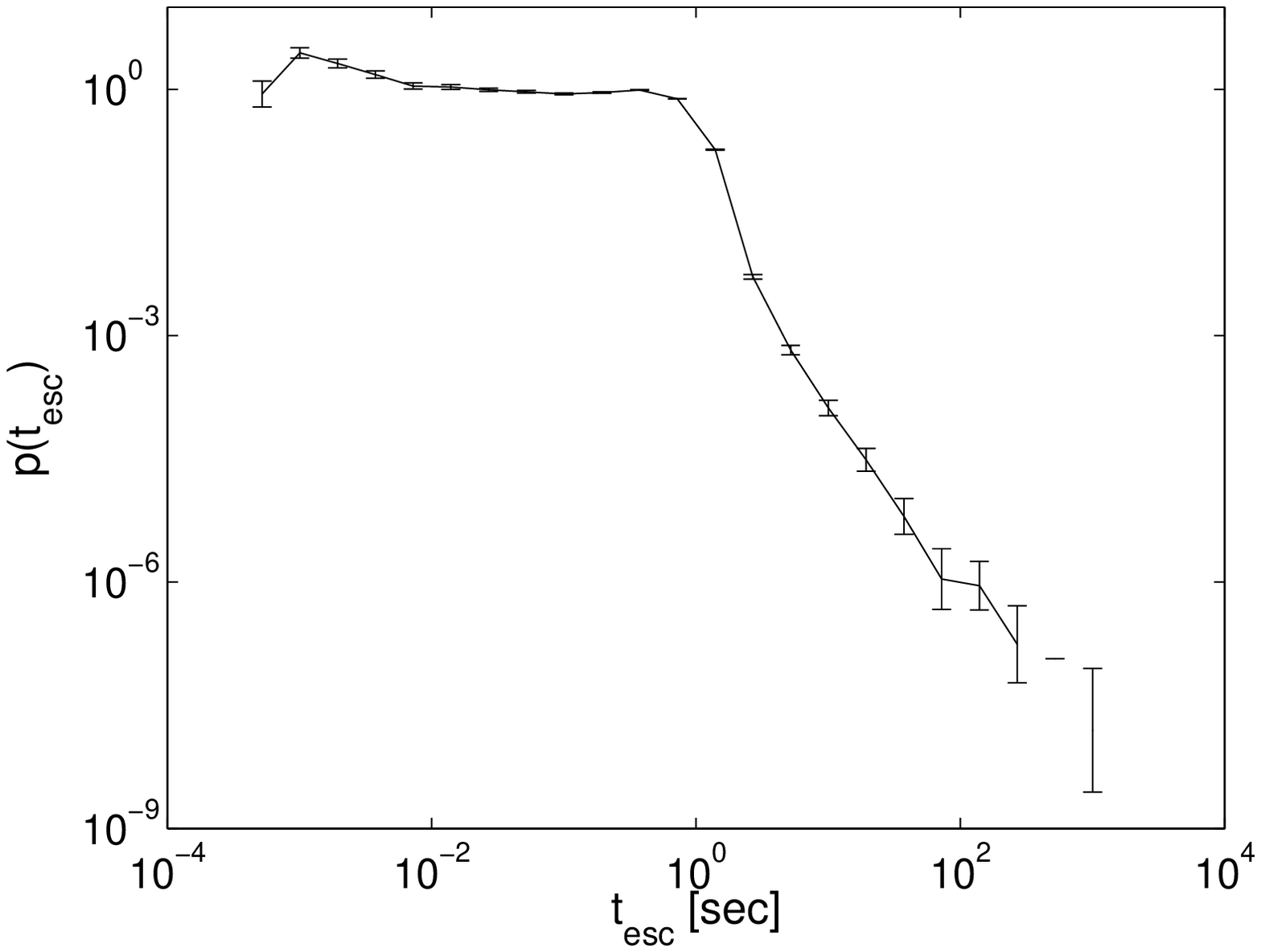}%
                  \label{f:distr_cs:tesc}}\\%
  \sidesubfloat[]{\includegraphics[width=0.43\columnwidth]{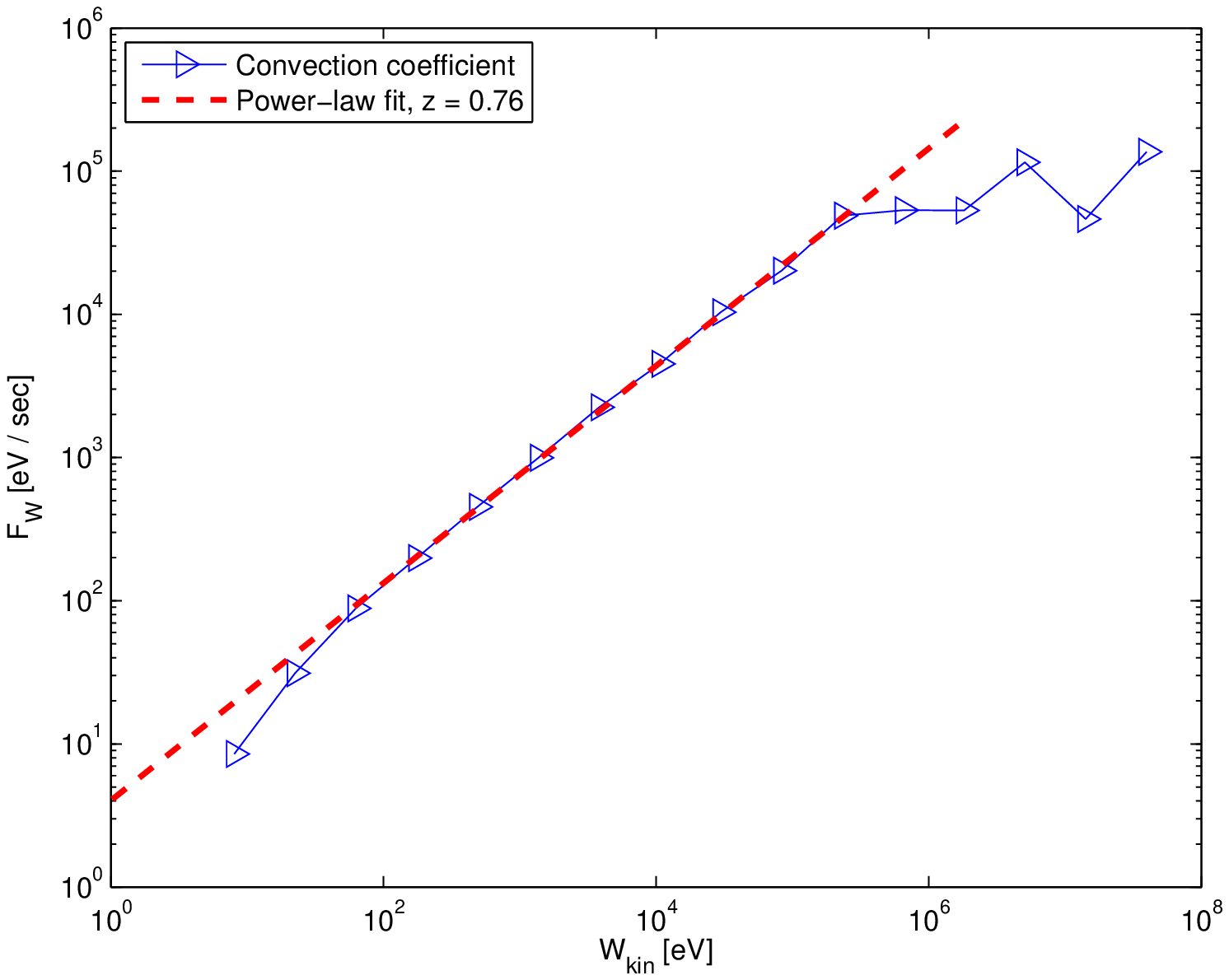}%
                  \label{f:distr_cs:F_W}}\hfill%
  \sidesubfloat[]{\includegraphics[width=0.43\columnwidth]{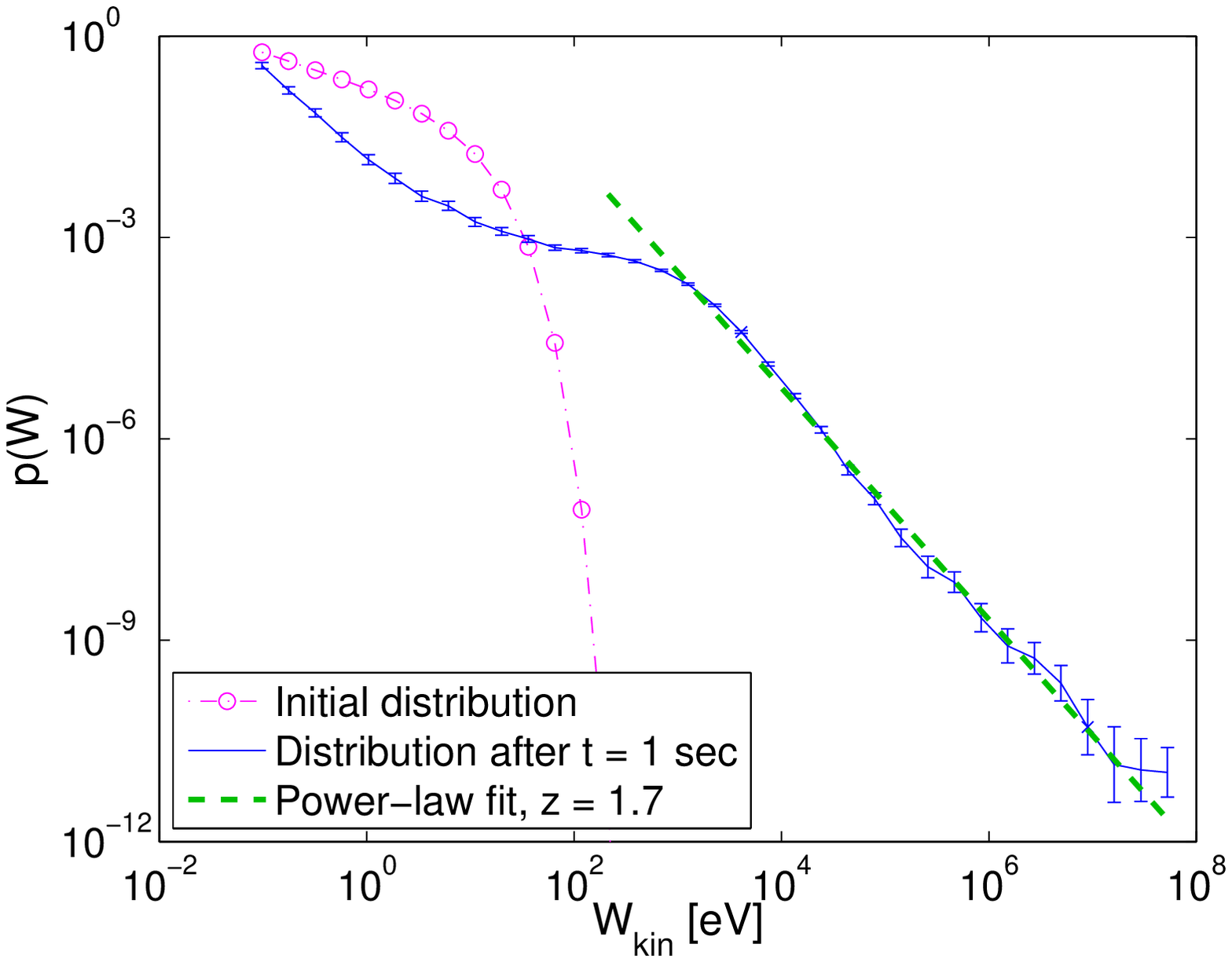}%
                  \label{f:distr_cs:Wdistr}}%
          \caption{\textit{\protect\subref{f:distr_cs:mW} The mean
  energy increase as a function of time and the energy evolution of
  some typical particles are shown.
  \protect\subref{f:distr_cs:tesc}  The escape time distribution of
  particles. \protect\subref{f:distr_cs:F_W} The energy convection
  coefficient as a function of the kinetic energy at $t = 1\ sec$.
  \protect\subref{f:distr_cs:Wdistr} Initial and asymptotic (here $t
  = 1 \ sec$) kinetic energy distribution.}}\label{f:distr_cs}
  \end{figure}

  We initiate the simulation with a Maxwellian distribution with
  temperature $10\ eV$. Fig.~(\ref{f:distr_cs:mW}) shows the mean energy and
  the energy of
  some  typical particles as a function of time, up to final time or
  untill they 
  escape from the simulation box. The rate at which the particles on the
  average gain
  energy is exponential, so $\ln\left< W \right> \approx t/t_{acc}$,
  and we estimate the asymptotic value of the acceleration time to be $\approx 0.3\; sec$. 
 
  The acceleration is systematic and the particles
  feel a rapid increase of their energy any time they cross an UCS
  with variable strength of the effective electric field (see the similar
  behaviour observed in \cite{Dahlin15,Guo15}). The energy
  distribution reaches an asymptotic state (see
  Fig.~(\ref{f:distr_cs:Wdistr})) in s fraction of a second. It is obvious that
  particles are very efficiently accelerated inside the turbulent
  reconnecting volume and form a power law tail with index $\approx
  1.7$. 

  Fig.~(\ref{f:distr_cs:tesc}) presents the escape time, which is
  different for each particle, and we use the median value $(\approx
  0.5\ sec)$ as an estimate of a characteristic escape time. If we
  use the estimates of $t_{acc}$ and $t_{esc}$ reported, we can
  estimate the index of the power law tail
  $k=1+t_{acc}/t_{esc} \approx 1.6$, which is  close  to the slope of the
  distribution of the accelerated particles in the simulation.

  In Fig.~(\ref{f:distr_cs:F_W}) the convection coefficient $F$ at $t
  = 1\ sec$ is presented as function of the energy, and it exhibits
  a power-law shape, with index $a_F = 0.76$ for energies above
  $100\,$eV, an index close to the one found above in Fermi's
  original scenario. For the diffusion coefficient, the estimate $D$
  based on Eq.~(\ref{eq:DWW}) yields a power-law, applying though
  the finite time correction of \cite{Ragwitz2001}, $D_{true} = D -
  0.5 \Delta t F^2$, we find that $D_{true} \approx 0$, the
  energization process is purely convective in nature, the non-zero
  $D$ is an artifact resulting from the finite time contribution of
  $F$ to $D$ (we just note that in the Fermi case the finite
  time correction
  was negligible).

  Using $F$ and $D_{true}$ in the numerical solution of the FP
  equation, we find only heating, on time-scales though of the order
  of tens of seconds, much larger than the time of 1 sec considered
  here. This result is in accordance with and a generalization of
  the result in \cite{Guo2014, Guo15}, who also find only heating
  when analytically solving the FP equation (for $D=0$ and $F\sim W$
  in their case). On the other hand, the asymptotic distribution can
  be calculated from Eq.~(\ref{diff}) (assuming $\partial n/\partial
  t =0$) as $n\sim W^{-0.76}$. The reason for the discrepancy
  between the FP solution and the asymptotic solution must be
  attributed to the fact that the asymptotic solution, determined as
  a stationary solution, cannot be reached with the initial
  condition being a Maxwellian (in analogy to the case in
  \cite{Guo2014} with $F\sim W$).

  Concerning the difference between the FP solution and the lattice
  model, we find that the sample of energy differences
  $W_i(t+\Delta t)- W_i(t)$ in Eq.~(\ref{eq:FW})
  (with $i$ the particle index),
  on which the estimate of $F$ is
  based, follows actually a power law distribution, and as a
  consequence the particles occasionally perform very large jumps in
  energy space (Levy flights), as illustrated in
  Fig.~(\ref{f:distr_cs:mW}), in contrast to the second order Fermi
  process (see Fig.~(\ref{f:distr_sof:mW})). The fact that the energy
  increments have a power-law distribution with the specific index
  has several consequences: (1) The estimate of $F$ as a mean  value
  theoretically is finite, yet it is very noisy. (2) Both the mean
  (or the median, as used here) are not representative of a
  scale-free power-law distribution. (3) The variance of the
  distribution of energy-increments tends to infinity. After all, in
  the case at hand, the applicability of the classical random walk
  theory (classical Langevin and FP equation) breaks down, as it is
  manifested in the inability of the FP equation to reproduce the
  test-particles' energy distribution, and in the practical
  difficulties of the expressions for $F$ and $D$ in
  Eqs.~(\ref{eq:FW}) and (\ref{eq:DWW}) to yield meaningful
  transport coefficients. Thus, modeling tools like the Fractional
  FP equation become appropriate here. Similar cases of Levy flights
  have been observed by \cite{Arzner04} and \cite{Bian08}, without
  further analyzing the consequences for the transport coefficients
  and the FP equation.

  We also have explored the role of collisions and they are
  important for impulsive energization longer than the collision
  time of the system,  they though play a crucial role only for the
  bulk of the energized plasma and just slightly modify the slope of
  the tail.
  %-----------------------------------------------------------------------------------

\section{Summary and Discussion}

 Turbulent reconnection is a new type of accelerator which can be modelled with the use of tools borrowed from Fermi type accelerators, namely by replacing the ``magnetic clouds" with  a new type of ``scatterers'', the UCS. This generalization can handle large scale astrophysical systems composed from local accelerators like current sheets appearing randomly in reconnecting turbulence. We developed a 2D lattice gas model where a number of active points act as ``scatterers" in order to model the new accelerator. Our main contribution in this article is the estimate of the transport coefficients from the particle dynamics and their use in solving the FP equation. Our  main results from this study are: (a) Stochastic Fermi accelerators can reproduce a well known energy distribution in laboratory and astrophysical plasmas, where heating of the bulk and acceleration of the run away tail co-exist. The density of the  scatterers plays a crucial role in controlling the heating and the acceleration of particles.  (b) The transport coefficients show a general power-law scaling
 	with energy.
 (c) The replacement of the scatterers with UCS has several effects on the energization of the particles: (i) The acceleration time is an order of magnitude faster than in the stochastic Fermi process. (ii) Estimating the transport coefficients from the dynamic particle orbits, we have shown that the final energy distribution cannot be a solution of the FP equation, since the orbits of the energetic particles in energy space depart radically from
 Brownian motion, showing characteristics of Levy flights. (iii) The asymptotic distribution of the accelerated particles  is similar to the ones obtained in  different simulations (see \cite{Arzner04,Dmitruk04,Onofri06, Drake06, Drake13,Dahlin15}),  where  turbulent reconnection is established.

We can conclude that the stochastic  Fermi acceleration and turbulent reconnection processes can play a crucial role in many astrophysical plasmas and their role depends strongly on their physical properties, such as
the nature of the scatterers (e.g.\ large amplitude Alfv\'en waves or UCS), their spatio-temporal statistical properties (e.g.\ their spatial density),
and the time evolution of the driver of the explosions.

\begin{acknowledgements}
We thank the referee, whose comments helped to improve substantily the article.
The authors acknowledge support by European Union (European Social Fund -ESF) and Greek national funds through the Operational Program Education and Lifelong Learning of the National Strategic Reference Framework (NSRF) -Research Funding Program: THALES: Investing in knowledge society through the European Social Fund.
\end{acknowledgements}

\bibliographystyle{aasjournal}
%\bibliography{vlahosastro}

\begin{thebibliography}{}
\expandafter\ifx\csname natexlab\endcsname\relax\def\natexlab#1{#1}\fi

\bibitem[{Arzner \& Vlahos(2004)}]{Arzner04}
Arzner, K., \& Vlahos, L. 2004, The Astrophysical Journal Letters, 605, L69

\bibitem[{Bian \& Browning(2008)}]{Bian08}
Bian, N.~H., \& Browning, P.~K. 2008, The Astrophysical Journal Letters, 687,
  L111

\bibitem[{{Biskamp} \& {Welter}(1989)}]{Biskamp89}
{Biskamp}, D., \& {Welter}, H. 1989, Physics of Fluids B, 1, 1964

\bibitem[{Boyd(2001)}]{Boyd2001}
Boyd, J.~P. 2001, Chebyshev and {{Fourier}} spectral methods ({Courier
  Corporation})

\bibitem[{Cargill {et~al.}(2012)Cargill, Vlahos, Baumann, Drake, \&
  Nordlund}]{Cargill12}
Cargill, P., Vlahos, L., Baumann, G., Drake, J., \& Nordlund, {\AA}. 2012,
  Space science reviews, 173, 223

\bibitem[{{Dahlin} {et~al.}(2015){Dahlin}, {Drake}, \& {Swisdak}}]{Dahlin15}
{Dahlin}, J.~T., {Drake}, J.~F., \& {Swisdak}, M. 2015, Physics of Plasmas, 22,
  100704

\bibitem[{{Dmitruk} {et~al.}(2004){Dmitruk}, {Matthaeus}, \&
  {Seenu}}]{Dmitruk04}
{Dmitruk}, P., {Matthaeus}, W.~H., \& {Seenu}, N. 2004, \apj, 617, 667

\bibitem[{{Drake} {et~al.}(2006){Drake}, {Swisdak}, {Che}, \& {Shay}}]{Drake06}
{Drake}, J.~F., {Swisdak}, M., {Che}, H., \& {Shay}, M.~A. 2006, \nat, 443, 553

\bibitem[{{Drake} {et~al.}(2013){Drake}, {Swisdak}, \& {Fermo}}]{Drake13}
{Drake}, J.~F., {Swisdak}, M., \& {Fermo}, R. 2013, The Astrophysical Journal
  Letters, 763, L5

\bibitem[{{Fermi}(1949)}]{Fermi49}
{Fermi}, E. 1949, Physical Review, 75, 1169

\bibitem[{{Galsgaard} \& {Nordlund}(1996)}]{Galsgaard96}
{Galsgaard}, K., \& {Nordlund}, {\AA}. 1996, Journal of Geophysical Research,
  101, 13445

\bibitem[{{Guo} {et~al.}(2014){Guo}, {Li}, {Daughton}, \& {Liu}}]{Guo2014}
{Guo}, F., {Li}, H., {Daughton}, W., \& {Liu}, Y.-H. 2014, Physical Review
  Letters, 113, 155005

\bibitem[{{Guo} {et~al.}(2015){Guo}, {Liu}, {Daughton}, \& {Li}}]{Guo15}
{Guo}, F., {Liu}, Y.-H., {Daughton}, W., \& {Li}, H. 2015, \apj, 806, 167

\bibitem[{Kulsrud \& Ferrari(1971)}]{Kulsrud71}
Kulsrud, R.~M., \& Ferrari, A. 1971, Astrophysics and Space Science, 12, 302

\bibitem[{{Lazarian} \& {Vishniac}(1999)}]{Lazarian99}
{Lazarian}, A., \& {Vishniac}, E.~T. 1999, \apj, 517, 700

\bibitem[{Lazarian {et~al.}(2012)Lazarian, Vlahos, Kowal, Yan, Beresnyak, \&
  Dal~Pino}]{Lazarian12}
Lazarian, A., Vlahos, L., Kowal, G., {et~al.} 2012, Space Science Reviews, 173,
  557

\bibitem[{{Longair}(2011)}]{Longair11}
{Longair}, M.~S. 2011, {High Energy Astrophysics}

\bibitem[{{Matthaeus} \& {Lamkin}(1986)}]{Matthaeus86}
{Matthaeus}, W.~H., \& {Lamkin}, S.~L. 1986, Physics of Fluids, 29, 2513

\bibitem[{{Onofri} {et~al.}(2006){Onofri}, {Isliker}, \& {Vlahos}}]{Onofri06}
{Onofri}, M., {Isliker}, H., \& {Vlahos}, L. 2006, Physical Review Letters, 96,
  151102

\bibitem[{Ragwitz \& Kantz(2001)}]{Ragwitz2001}
Ragwitz, M., \& Kantz, H. 2001, Physical Review Letters, 87, 254501

\end{thebibliography}

% =======================================
\end{document}